\documentclass[aps,prd,showpacs]{revtex4}
\usepackage{graphicx}
\begin{document}
\begin{flushright}{OITS 709}\\
February 2002
\end{flushright}

\vspace*{1cm}

\title{Parton Distributions in the Valon Model}
\author{\bf   Rudolph C. Hwa$^1$ and  C.\ B.\ Yang$^{1,2}$}
\affiliation{$^1$Institute of Theoretical Science and Department of Physics\\
University of Oregon, Eugene, OR 97403-5203, USA\\
\bigskip
$^2$Institute of Particle Physics, Hua-Zhong Normal University,
Wuhan 430079, P.\ R.\ China}

\vskip.5cm
\begin{abstract}
The parton distribution functions determined by CTEQ at low
$Q^2$ are used as inputs to test the validity of the valon model.
The valon distributions in a nucleon are first found to be nearly
$Q$ independent.  The parton distribution in a valon are shown
to be consistent with being universal, independent of the valon
type. The momentum fractions of the partons in the valon add up
separately to one.  These properties affirm the validity of the
valon model.  The various distributions are parameterized for
convenient application of  the model.

\end{abstract}
\maketitle

\section{Introduction}

The parton distributions in proton have been studied extensively in
recent years over wide ranges of $Q^2$ and $x$
\cite{ct,mrs,grv}.  For example, in the CTEQ global analysis
framework
\cite{ct} the distribution functions have been determined by fitting
some 1300 data points obtained for various reactions in 16
experiments; over 30 parameters are used in the input parton
distribution functions (PDF's) in the next-to-leading -order
calculations in perturbative QCD \cite{gs}. The results on the PDF's at
various $Q^2$ are presented in the form of graphs available on the
web \cite{ct4}. The distribution functions are accurately calculated
numerically, but they are inconvenient to describe analytically. The
purpose of this paper is twofold. One is to provide a simple
parameterization of the parton distribution functions that is
accurate to within 5\% for $1<Q^2<100$ (GeV/c)$^2$ and
$10^{-2}<x<1$. The other is to provide firm evidence for the
validity of a model that can be very useful in the study of soft
processes in hadronic and nuclear collisions.

The model under discussion is the valon model \cite{hvm,hz}.
Valons play a role in scattering problems as the constituent quarks
do in bound-state problems. In the model it is assumed that the
valons stand at a level in between hadrons and partons and that the
structure of a hadron in terms of the valons is independent of
$Q^2$. That is, the property that a nucleon has three valons which
carry all the momentum of the nucleon does not change with
$Q^2$. Each valon may be viewed as a parton cluster associated with
one and only one valence quark, so the flavor quantum
numbers of a valon are those of a valence quark. The $Q^2$
evolution of the parton distributions in a nucleon is effected through
the evolution of the valon structure, as the higher resolution of a
probe reveals the parton content of the valons.

When the valon model was first proposed, the deep inelastic
scattering data was not precise enough either to rule out the model
or to determine accurately the parameters in the model. One may
regard the situation then as one in which the model satisfies the
sufficiency condition for an approximate description of the nucleon
structure, but not necessary. Now, the experimental data have vastly
improved and the PDF's have been so precisely determined that the
validity of the valon model can be put to a stringent test. That is
what we intend to do in this paper. Although no model is ever
necessary in the mathematical sense, we shall show that the concept
of valons as constituents of the nucleon is eminently acceptable by
virtue of the $Q^2$ independence of the valon distribution
functions, the universality of the parton distributions in valons,
and the momentum sum rule of the partons in a valon being
satisfied.

In terms of the parameterization in the valon model the parton
distributions are much simpler to describe and therefore more
convenient to transport to different problems. Our focus will be on
soft processes at low $Q^2$, for which perturbative QCD is
inapplicable. The valon picture then provides a systematic way of
organizing various contributions to inclusive processes. One such
problem is the study of proton-nucleus collisions in which the
degradation of the momenta of the produced nucleons can be well
described in the valon model \cite{hy}. The new parameterizations
determined in this paper will affect the details of the model, for
which we have previously made simplifying assumptions. The
application of the results obtained now will not be considered here.
We mention such applications here only to motivate our
concentration in the range $1<Q^2<100$ (GeV/c)$^2$ in this
paper.

\section {The Valon Model}

In the valon model we assume that a proton consists of three valons
$(UUD)$, which separately contain the three valence quarks
$(uud)$. Let the exclusive valon distribution function be
\begin{eqnarray}
G_{UUD}(y_1, y_2, y_3) = g \, (y_1y_2)^{\alpha}y^{\beta}_3
\, \delta (y_1 + y_2 + y_3 -1) ,
\label{2.1}
\end{eqnarray}
where $y_i$ are the momentum fractions of the $U$ valons
$(i=1, 2)$ and $D$ valon $(i=3)$. The variable $y$ will  never refer
to rapidity in this paper. The normalization factor $g$ is determined
by the requirement that the probability for the proton to consist of
three and only three valons is one, i.e.,
\begin{eqnarray}
\int^1_0 dy_1 \int^{1 - y_1}_0dy_2 \int^{1 - y_1-y_2}_0dy_3 \,
G_{UUD}(y_1, y_2, y_3) = 1 .
\label{2.2}
\end{eqnarray}
Note that the valon distribution function is not defined in the
invariant phase space.  From Eq.\ (\ref{2.2}) we have
\begin{eqnarray}
g = \left[B (\alpha + 1, \beta +1) B(\alpha + 1,
\alpha +\beta +2)\right]^{-1},
\label{2.3}
\end{eqnarray}
where $B(m,n)$ is the beta function. The single-valon distributions
are
\begin{eqnarray}
G_U(y) = \int^{1-y}_0 dy_2 \int^{1-y-y_2}_0
dy_3 G_{UUD} (y, y_2, y_3) = g B (\alpha + 1,
\beta +1) y^{\alpha} (1 - y)^{\alpha +\beta +1},
\label{2.4}
\end{eqnarray}
\begin{eqnarray}
G_D(y) = \int^{1-y}_0 dy_1 \int^{1-y-y_1}_0
dy_2 G_{UUD} (y_1, y_2, y)= g B (\alpha + 1, \alpha +1) y^{\beta}
(1 - y)^{2\alpha +1}.
\label{2.5}
\end{eqnarray}

An essential property of the valon model is that the structure of the
proton in terms of the valons is independent of the probe. It means
that when probed at high $Q^2$, whatever the experiment may be,
the parton distributions in a proton can be expressed as a
convolution of the valon distribution and the parton distribution in
a valon, i.e.,
\begin{eqnarray} x \, u (x,Q^2) = \int^1_x dy \left[ 2G_U (y)
K(x/y,Q^2)
 + G_D(y) L_u (x/y,Q^2)\right],
\label{2.6}
\end{eqnarray}
\begin{eqnarray} x \, d (x,Q^2) = \int^1_x dy \left[G_D (y)
K(x/y,Q^2)
 + 2 G_U(y) L_u (x/y,Q^2)\right],
\label{2.7}
\end{eqnarray}
where $u(x, Q^2)$ and $d(x,
Q^2)$ are the $u$ and $d$ quark distributions, respectively, and
$x$ the momentum fraction of the quark. We emphasize that on the
right-hand side (RHS) of the above equations the $Q^2$
dependences appear only in parton distributions in the valons, $K(z,
Q^2)$ and $L_u(z, Q^2)$, but not in the valon distributions,
$G_U(y)$ and $G_D(y)$. We regard this as the defining property of
the valon model. In that sense the model is analogous to the one in
which a deuteron is treated as a bound state of two nucleons; in that
treatment a high-$Q^2$ probe resolves the structure of one or the
other nucleon without affecting the nucleon wave function of the
deuteron.

There are two types of parton distributions in the valons that
appear in Eqs. (\ref{2.6}) and (\ref{2.7}).  $K(z, Q^2)$ refers to the
favored partons, i.e., $u$ in $U$ and $d$ in $D$, whereas $L_u(z,
Q^2)$ refers to the unfavored partons, i.e., $u$ in $D$ and $d$ in
$U$.  The distribution $K (z, Q^2)$ can be further divided into two
types
\begin{eqnarray} K(z,Q^2 ) = K_{NS}(z,Q^2 ) + L_f (z,Q^2 ),
\label{2.8}
\end{eqnarray}
where the first term on the RHS is the valence quark
distribution (hence, non-singlet), while the second is the sea quark
distribution for the same flavor type. Since the sea quarks should
respect charge conjugation invariance, the $u$ and $\bar{u}$ in the
sea (and similarly $d$ and $\bar{d}$) have the same distributions,
i.e.,
\begin{eqnarray}
x \, \bar{u} (x, Q^2) = \int^1_x dy \left[2G_U (y)
L_f (x/y, Q^2)
 + G_D(y) L_u (x/y, Q^2)\right],
\label{2.8a}
\end{eqnarray}
\begin{eqnarray}
x \, \bar{d} (x, Q^2) = \int^1_x dy \left[G_D (y)
L_f (x/y, Q^2)
 + 2 G_U(y) L_u (x/y, Q^2)\right].
\label{2.9}
\end{eqnarray}
The valence quark distributions are then
\begin{eqnarray} x \, u_v (x,Q^2) =  \int^1_x dy 2 G_U(y) K_{NS}
(x/y,Q^2),
\label{2.10}
\end{eqnarray}
\begin{eqnarray} x \, d_v (x,Q^2) = \int^1_x dy G_D (y) K_{NS}
(x/y,Q^2)  .
\label{2.11}
\end{eqnarray}

In earlier treatment \cite{hvm}-\cite{hy} $L_f$ and $L_u$ have been
regarded as identical due to the assumption of the symmetric sea.
Indeed, putting $L_f = L_u$ in Eqs. (\ref{2.8a}) and (\ref{2.9})
results in $\bar{u}(x, Q^2) = \bar{d}(x, Q^2)$, which is a necessary
consequence of the sea quarks satisfying $SU(2)$ symmetry.
However, there is experimental evidence \cite{ma} that Gottfried
integral $\int (F_2^p-F_2^n) dx/x$ is less than $1/3$, which is the
value expected in the simple quark model. Thus we should allow
$L_f$ to be different from $L_u$. Indeed, in the valon model we may
expect that in a $U$ valon the necessary presence of a $u$ valence
quark would on the grounds of Fermi statistics make a gluon have
more difficulty converting virtually into a $u\bar{u}$ pair than into
a $d\bar{d}$ pair. Hence, $L_f$ should be suppressed relative to
$L_u$. Since there are two $U$ and one $D$ is a proton, we should
expect the overall
$\bar{u}$ to be less than $\bar{d}$. The data do indicate
$\bar{u}<\bar{d}$ after integration over $x$ \cite{ma}.  We thus see
that the breaking of
$SU(2)$ in the sea is related to Pauli blocking in the valons.

\section{The Valon Distributions}

    The valence quark distributions, $u_v$ and $d_v$, are given by
CTEQ4LQ \cite{ct4}.  From Eqs.\ (\ref{2.10}), and (\ref{2.11}) we see
that they are directly related to the valon distributions, $G_U(y)$
and $G_D(y)$, by convolutions with the common factor $K_{NS}$. It
is therefore possible to isolate the valon distributions by
deconvolution using the moments. Let us define
\begin{eqnarray}
\tilde{G}_{U, D} (n) = \int^1_0 dy y^{n-1} G_{U, D} (y),
\label{3.1}
\end{eqnarray}
\begin{eqnarray}
\tilde{K}_{NS} (n,Q^2)  = \int^1_0 dz z^{n-2}  K_{NS}
(z,Q^2),
\label{3.2}
\end{eqnarray}
\begin{eqnarray}
\tilde{u}_v  (n,Q^2) = \int^1_0 dx x^{n-1} u_v (x,Q^2),
\label{3.3}
\end{eqnarray}
and similarly for $\tilde{d}_v$ in terms of $d_v$. Then by the
convolution theorem we have from Eqs. (\ref{2.10}) and
(\ref{2.11}),
\begin{eqnarray}
\tilde{u}_v  (n,Q^2) =  2\tilde{G}_U(n) \tilde{K}_{NS} (n,Q^2),
\label{3.4}
\end{eqnarray}
\begin{eqnarray}
\tilde{d}_v  (n,Q^2) =  \tilde{G}_D(n) \tilde{K}_{NS} (n,Q^2).
\label{3.5}
\end{eqnarray}
It thus follows that
\begin{eqnarray}
{\tilde{G}_U(n) \over \tilde{G}_D(n)} =
{\tilde{u}_v  (n,Q^2)
\over 2 \tilde{d}_v  (n,Q^2)} .
\label{3.6}
\end{eqnarray}
If the valon model is valid, then the LHS is
$Q^2$ independent, a property that we can check directly by
examining the
$Q^2$ dependence of the RHS. Since $u_v(x, Q^2)$ and  $d_v(x,
Q^2)$ can separately be determined from \cite{ct4}, we only have to
take their moments and calculate their ratio.

To do the above analysis, we have to set the range of $Q^2$ to be
examined, since the valon model is not expected to be accurate for
all
$Q^2$. As we have discussed near the end of Sec. I, the valon model
has been applied to soft production problems because they involve
non-perturbative processes. For hard processes at very high
$Q^2 \, [Q^2 >100$ (GeV/c)$^2]$ perturbative QCD is very
successful and there is no need to introduce any inaccuracies through
the use of a model. We shall therefore limit ourselves to the range
$1<Q^2<100 $(GeV/c)$^2$. This is actually a very wide range for
hadronic processes that can involve the production of soft
particles and semi-hard mini-jets.

For the range of $Q^2$ chosen we must use low-$Q^2$
parameterization of the PDF's.  CTEQ4LQ \cite{ct4} gives the graphs
of $u$, $d$, $s$, $\bar{u}$, $\bar{d}$, and $g$ distributions at
any $Q$ evolved from the starting scale at $Q^2_0 =
0.49$~(GeV/c)$^2$.  Since $u_v$ and $d_v$ distributions are not
included in the list of PDF's posted, we have to calculate
\begin{eqnarray}
u_v (x,Q^2) = u (x,Q^2) - \bar{u} (x,Q^2)
\label{3.7}
\end{eqnarray}
\begin{eqnarray}
d_v (x,Q^2) = d (x,Q^2) - \bar{d} (x,Q^2)
\label{3.8}
\end{eqnarray}
from the $q$ and $\bar{q}$  graphs for the RHS available
from the web.  We extract the numerical values at up to 60 points of
$x$ values per PDF for three values of $Q:  1, 3, 10$ GeV/c.  From
the values of
$u_v$ and
$d_v$ thus determined, we then compute the moments in
accordance to Eq.\ (\ref{3.3}) for $n = 2, \cdots 9$.  Denoting the
RHS of Eq.\ (\ref{3.6}) by $r(n, Q)$ we then calculate the ratio of
ratios
\begin{eqnarray}
R(n, Q) = r (n, Q)/r(n, Q = 1)
\label{3.9}
\end{eqnarray}
relative to $Q = 1$ GeV/c.  Figure 1 shows how $R(n, Q)$ depends on
$n$ for $Q = 3$ and 10 GeV/c.  It is evident that the dependence is
very mild, the maximum deviation from 1 being around 7\% at $n =
9$ and $Q = 10$ GeV/c.  We regard this approximate constancy of
$R(n, Q)$, while $Q$ is increased by an order of magnitude, as the
first step toward a confirmation of the validity of the valon model.
That is, the insensitivity of $R(n, Q)$ to $Q$ variation supports our
assumption that $\tilde{G}_U(n)/\tilde{G}_D(n)$ is $Q$
independent to a degree sufficient for the application of the valon
model.

The $n$ dependence of $r(n, Q) = \tilde{u}_v(n, Q)/2\tilde{d}_v(n,
Q)$, as determined from the CTEQ4LQ data, can now be used to fix
the parameters, $\alpha$ and $\beta$, in the valon distributions.
From Eqs.\ (\ref{2.4}), (\ref{2.5}) and (\ref{3.1}) we have
\begin{eqnarray}
\tilde{G}_U(n) = B (\alpha + n, \alpha + \beta + 2)/B(\alpha + 1,
\alpha + \beta +2)
\label{3.10}
\end{eqnarray}
\begin{eqnarray}
\tilde{G}_D(n) = B (\beta + n, 2 \alpha + 2)/B(\beta + 1,
2 \alpha + 2)
\label{3.11}
\end{eqnarray}
from which follows
\begin{eqnarray}
{\tilde{G}_U(n) \over  \tilde{G}_D(n)}= {\Gamma (\alpha +
n)\Gamma (\beta + 1)
\over  \Gamma (\alpha +
1)\Gamma (\beta + n)}
\equiv
\gamma(\alpha, \beta, n).
\label{3.12}
\end{eqnarray}
The parameters $\alpha$ and $\beta$ can now be determined by
minimizing
\begin{eqnarray}
C = \sum^N_{n=2} \left[{\gamma(\alpha, \beta, n) -  r(n,
Q) \over \gamma(\alpha, \beta, n) +  r(n,
Q)}\right]^2,
\label{3.13}
\end{eqnarray}
where $N$ is the maximum number of moments we extract from the
CTEQ4LQ data, which we take to be $N = 10$.  Note that $n = 1$
is excluded in Eq.\ (\ref{3.13}), since  $\gamma(\alpha, \beta, 1) = 1$
basically due to Eq.\ (\ref{2.2}), and $r(1, Q) = 1$ because there are
two $u_v$ and one
$d_v$.  Varying $\alpha$ and $\beta$ in search for a minimum in
$C$ results  in incredibly small $C$, in the order of $10^{-5}$.  We
find
\begin{eqnarray}
\alpha = 1.755, \qquad \beta = 1.05, \qquad \qquad\mbox{for} \  Q
= 1 \,
\mbox{GeV}/c .
\label{3.14}
\end{eqnarray}
\begin{eqnarray}
\alpha = 1.545, \qquad \beta = 0.89, \qquad \qquad \mbox{for} \  Q
= 10 \,
\mbox{GeV}/c,
\label{3.15}
\end{eqnarray}
We have not ignored the small $Q$ dependences of $\alpha$ and
$\beta$ since the data on $r(n, Q)$ contain some $Q$ dependence.
Moreover, the $Q$ independence of the ratio $\tilde{G}_U(n)/
\tilde{G}_D(n)$ does not mathematically preclude the $Q$
dependences of $\tilde{G}_U(n)$ and $\tilde{G}_D(n)$ separately.
However, the difference between Eqs.\ (\ref{3.14}) and (\ref{3.15})
is not very great, as we can see in Fig.\ 2, where $G_U(y)$ and
$G_D(y)$ are shown [through the use of Eqs.\ (\ref{2.4}) and
(\ref{2.5})] for the two extreme $Q$ values, $1$ and $10$ GeV/c.
The difference is insignificant compared to those of the quark
distributions, one of which is shown in Fig.\ 3.  With the drastic
difference between Figs.\ 2 and 3 in mind, it is reasonable to
conclude that the essence of the valon model has been verified to
the extent that the valon distributions exhibit approximate scaling
in $Q$.

It should be remarked that the values of $\alpha$ and $\beta$
determined above are very different from the ones obtained
previously.   In Ref.\ \cite{hz} the early data of muon \cite{bag} and
electron scattering \cite{abe} were used in conjunction with a
number of theoretical assumptions to yield the values of $\alpha =
0.65$ and $\beta = 0.35$.  In Ref.\
\cite{hy} the modern data of CTEQ4LQ were used, but the values
$\alpha = 0.70$ and $\beta = 0.25$ were obtained due to the
assumption of a specific form of the sea quark distribution (proven
to be grossly inaccurate below).  Here in this
paper we have avoided making any assumption about the sea quark
distributions.  By extracting the valence quark distributions from
CTEQ4LQ we have been able to determine $G_U(y)$ and
$G_D(y)$ directly.  Clearly, the new values of $\alpha$ and $\beta$
are more reliable.  Further support for their reliability will be given
below in connect with the quark distributions $u$ and $d$, for
which our previous parameterization in Ref.\ \cite{hy} has led to
unaccountable discrepancies that are unsatisfactory.

From Eq.\ (\ref{3.1}) one sees that the momentum fractions of the
valons are given by the $n=2$  moments.  From Eqs.\
(\ref{3.10}),  (\ref{3.11}), (\ref{3.14}) and (\ref{3.15}), one can
then calculate the momentum fractions $\left<y\right>$, yielding
\begin{eqnarray}
\left<y\right>_U = &0.3644, &\qquad\qquad Q = 1 \, \mbox{GeV}/c,
\nonumber
\\ &0.3646,&\qquad\qquad Q = 10 \, \mbox{GeV}/c,
\label{3.16}
\end{eqnarray}
\begin{eqnarray}
\left<y\right>_D = &0.2712 &\qquad\qquad Q = 1 \,
\mbox{GeV}/c,\nonumber
\\ &0.2708,&\qquad\qquad Q = 10 \,
\mbox{GeV}/c.
\label{3.17}
\end{eqnarray}
At either $Q$ the sum rule
\begin{eqnarray}
2\left<y\right>_U + \left<y\right>_D = 1
\label{3.18}
\end{eqnarray}
is satisfied identically. We see that the momentum fractions
carried by the valons are essentially independent of $Q$ and that
each $U$ valon carries as much as 1.345 times more than the $D$
valon.

\section{The Quark and Gluon Distributions}

Having determined the valon distributions, we can now proceed to the
quark and gluon distributions in the valons.  From Eq.\ (\ref{3.4}) we
have
\begin{eqnarray}
\tilde{K}_{NS} (n, Q) = \tilde{u}_v (n, Q)/2 \tilde{G}_U(n,Q)
\label{4.1}
\end{eqnarray}
where we allow $\tilde{G}_U(n,Q)$ to have its weak $Q$ dependence
given by Eqs.\ (\ref{3.14}) and (\ref{3.15}).  Using the moments
$\tilde{u}_v (n, Q)$ that we have already calculated from CTEQ4LQ,
we obtain the values of $\tilde{K}_{NS} (n, Q)$ as shown in Fig.\ 4.
As expected, $\tilde{K}_{NS} (n, Q)$ undergoes substantial evolution,
especially from $Q = 1$ to $3$ GeV/c.

To test how good our determination of $\alpha$ and $\beta$ is, we
use the $\tilde{K}_{NS} (n, Q)$ calculated above in conjunction with
$\tilde{G}_D(n,Q)$ that can be obtained from Eqs.\ (\ref{2.5}) and
(\ref{3.1}) so that $\tilde{d}_v (n, Q)$ can be computed using Eq.\
(\ref{3.5}).  Note that this computation of $\tilde{d}_v (n, Q)$
requires the knowledge of $\alpha$ and $\beta$, while the
computation of $\tilde{d}_v (n, Q)$ for the RHS of Eq.\ (\ref{3.6}),
i.e., $r(n, Q)$, is based on the CTEQ4LQ data for the RHS of Eq.\
(\ref{3.8}).  Our point here is to calculate $d(x,Q)$ from the CTEQ
input on $\tilde{u}_v (n, Q)$ in Eq.\ (\ref{4.1}), a procedure that is
made possible by the common factor $\tilde{K}_{NS} (n, Q)$ in both
Eqs.\ (\ref{3.4}) and (\ref{3.5}).  Physically, it means that the
evolution of quarks in a valon is independent of the flavor of the host
valon.

After $\tilde{d}_v (n, Q)$ is obtained by the above procedure that
tests the values of $\alpha$ and $\beta$, we then make the inverse
transform to get the distribution $d_v(x, Q)$.  This transform can be
facilitated by exploiting the orthogonality of the Legendre
polynomials, the details of which are discussed in Ref.\ \cite{hy}.
Upon the determination of $d_v(x, Q)$ we can add to it the
$\bar{d}(x, Q)$ distribution from Ref.\ \cite{ct4} and obtain $d(x,
Q)$.  In Fig. 5 we show the $u$ and $d$ quark distributions at $Q = 1$
GeV/c.  The solid lines are the distributions posted by CTEQ4LQ
\cite{ct4}.  The dotted line for $xd(x)$ is what we have computed
using the procedure outlined above.  Note that its agreement with the
solid line is excellent.  The dotted line for $xu(x)$ is essentially the
result from fitting the CTEQ4LQ data in the valon model; it merely
affirms that the fit is extremely good, so the values of $\alpha$ and
$\beta$ are reliable.  The result on $xd(x)$ reveals more about the
soundness of the model, since it is not obtained by fitting, but
calculated using the valon distribution $G_D(y)$ and the universality
of $K_{NS}(z, Q)$.

For the sea quark and gluon distributions, it is for the convenience of
the applications of the valon model that we find simple
parameterizations of their distributions in a valon.  To that end we
first write Eqs.\ (\ref{2.8a}) and (\ref{2.9}) in moment form
\begin{eqnarray}
\tilde{\bar{u}} (n, Q) =  2\tilde{G}_U (n)
\tilde{L}_f (n, Q)
 + \tilde{G}_D(n) \tilde{L}_u (n, Q),
\label{4.2}
\end{eqnarray}
\begin{eqnarray}
\tilde{\bar{d}} (n, Q) = \tilde{G}_D(n) \tilde{L}_f (n, Q) +
 2\tilde{G}_U (n) \tilde{L}_u (n, Q) .
\label{4.3}
\end{eqnarray}
For the strange quark and gluons, we have
\begin{eqnarray}
\tilde{s} (n, Q) =  \left[2\tilde{G}_U (n)
 + \tilde{G}_D(n)\right] \tilde{L}_s (n, Q),
\label{4.4}
\end{eqnarray}
\begin{eqnarray}
\tilde{g} (n, Q) = \left[2\tilde{G}_U (n)
 + \tilde{G}_D(n)\right] \tilde{L}_g (n, Q) .
\label{4.5}
\end{eqnarray}
Since $\tilde{\bar{u}}$, $\tilde{\bar{d}}$, $\tilde{s}$ and $\tilde{g}$
are known from CTEQ4LQ, and $\tilde{G}_U$ and $\tilde{G}_D$
known from Eqs.\ (\ref{3.10}) and (\ref{3.11}), we can solve for
$\tilde{L}_f$, $\tilde{L}_u$, $\tilde{L}_s$ and $\tilde{L}_g$.  From
the result we perform the inverse transform to $L_f (z, Q)$, $L_u (z,
Q)$, $L_s (z, Q)$ and $L_g (z, Q)$, respectively.  These distributions
are shown in Figs. 6 and 7, for $Q = 1, 3,$ and $10$ GeV/c.  Being in
log-log plots, the evolutions due to the changes in $Q$ are
substantial, as expected.  What we have not expected is the drastic
difference between $L_f (z, Q)$ and $L_u (z, Q)$.  For $0 < -\ln (1 - z)
< 0.4$, the range of $z$ is
$0 < z < 0.33$.  For $z > 0.4$ we find that $L_f (z, Q) \ll L_u (z,
Q)$, at least for $Q = 1$ and $3$ GeV/c.  At higher $Q$ the evolution
can generate more favored sea quarks.  Thus at low $Q$, where sea
quarks are few, Pauli blocking suppresses the favored sea quarks so
much at high $z$ that the unfavored sea quarks dominate.  Indeed,
our calculation of $L_f (z, Q)$ is unreliable for $z > 0.4$ because
from the finite number of moments $(n < 10)$ that we have taken the
inverse transform generates oscillations in $z$ at high $z$.  There is
no such problem with the $s$ quark and gluon distributions, as is
evident in Fig. 7, since they are not inhibited by Pauli blocking.  The
general property is that all the parton distributions increase
significantly at small $z$, when $Q$ is increased.

There is no simple way to describe both the $z$ and $Q$
dependences of the parton distributions.  In order for the valon model
to be useful, especially in applications to the study of inclusive cross
sections in hadronic collisions at low $p_T$, an analytic description
of each of the parton distributions is needed.  For such problems only
the distributions at $Q = 1$ GeV/c are relevant, so we fit those
distributions by polynomials.  The formula used for the fitting is
\begin{eqnarray}
\ln L_i (z, Q = 1) = \sum^3_{j = 0} a^{(i)}_j t^j, \qquad \qquad t =
- \ln (1 - z).
\label{4.6}
\end{eqnarray}
The result of the fitting is shown in Fig. 8, for which the values of the
coefficients are given in Table I.  The fits are evidently very good.
Thus we have completely specified the PDF's at $Q = 1$ GeV/c in
analytical and numerical ways, suitable for transport to problems
where such PDF's are needed.
\begin{table}[ht]
\begin{center}
\caption{Coefficients in Eq.\ (\ref{4.6})}
\begin{tabular}{|c|c|c|c|c|}\hline
$i$&$a^{(i)}_0$&$a^{(i)}_1$&$a^{(i)}_2$&$a^{(i)}_3$\\\hline
$f$&-2.66&0.08&-10.4&-6.0\\\hline
$u$&-2.92&4.0&-5.95&-1.4\\\hline
$s$&-3.30&-2.4&2.7&-1.65\\\hline
$g$&-0.63&-3.1&4.9&-1.9\\\hline\hline
\end{tabular}
\end{center}
\end{table}

In applications, such as in Ref.\ \cite{hy}, the moments are more
useful than the PDF's themselves.  We show the moments at $Q =
1$ GeV/c in Fig.\ 9, together with their fits.  The fitting
is done mainly for the convenience of applications. The formula
used for fitting is
\begin{eqnarray}
\ln \tilde{L}_i(n) = -\sum^3_{j = 0} b^{(i)}_j u^j, \qquad\qquad
 u = \ln (n -1),
\label{4.7}
\end{eqnarray}
 where the coefficients are given in
Table II.
\begin{table}[ht]
\begin{center}
\caption{Coefficients in Eq.\ (\ref{4.7})}
\begin{tabular}{|c|c|c|c|c|}\hline
$i$&$b^{(i)}_0$&$b^{(i)}_1$&$b^{(i)}_2$&$b^{(i)}_3$\\\hline
$f$&4.12&2.2&0.2&0.18\\\hline
$u$&3.07&1.5&0.08&0.05\\\hline
$s$&4.21&1.6&0.1&0.02\\\hline
$g$&0.98&1.0&0.05&0\\\hline\hline
\end{tabular}
\end{center}
\end{table}
The rapid decrease of the favored quark moments
$\tilde{L}_f(n)$ with increasing $n$ is now very evident, while
the other three parton moments have roughly similar $n$
dependences.

It is important to check the momentum sum rule of the partons.
We have seen in Eq.\  (\ref{3.18}) that the valon momentum
fractions add up to 1;  now the parton momentum fractions in
each valon must also add up to 1. Since $K_{NS}(z)$ and $L_i(z)$
are invariant distributions, their moments at $n=2$ are their
momentum fractions. We therefore should have
\begin{eqnarray}
\tilde{K}_{NS} (2) + \tilde{L}_g(2) + 2 \sum _{i = f, u. s}
\tilde{L}_i(2) = 1 .
\label{4.8}
\end{eqnarray}
We have $\tilde{K}_{NS}(2) = 0.4707$ at $Q=1$ GeV/c, and from
Table II we can calculate $\tilde{L}_i(2) = \exp [-b^{(i)}_0]$, yielding
$0.0162$, $0.0465$, $0.0148$, and $0.3754$ for $i=f$, $u$, $s$,
and $g$, respectively.  According to the LHS of Eq.\ (\ref{4.8})
they sum up to $1.001$, an excellent confirmation of the
momentum sum rule.

As a final item of paremetrizing the moments of the parton
distributions, we give here also a formula that fits
$\tilde{K}_{NS} (n, Q)$ at $Q=1$ GeV/c
\begin{eqnarray}
\ln \tilde{K}_{NS} (n) = - \sum^3_{j = 0} c_j u^j ,
\label{4.9}
\end{eqnarray}
where $c_j = 0.753$, $0.401$, $0.0962$, and $0.0555$, for $j =
0, 1, 2, 3,$ respectively.  The solid line in Fig. 4 shows the fit.

\section {Conclusion}

We have shown that the PDF's in a proton can be described in two
stages:  valons in a proton and then partons in a valon. The valon
distribution functions are essentially independent of $Q$, while the
parton distributions in the valons are $Q$ dependent. The three
valons carry all the momentum of the proton, and the way that
the parton momenta are distributed in a valon is independent of
the host valon so long as the sea quark flavors are identified as
favored or unfavored, instead of by specific flavors like $u$ or
$d$. We have found that Pauli blocking significantly suppress the
favored quarks compared to the unfavored quarks. At $Q=1$
GeV/c, the valence quarks carry 47.1\% of the proton
momentum, while the gluons carry 37.5\%. The scaling behavior
of the valon distributions and the universality of the valon
structure give support to the valon model as a simple and
organized description of the nucleon structure.

We have determined the parameters in simple formulas that
adequately describe the parton distributions (and their moments)
in a valon at $Q=1$ GeV. This is done for the benefit of
applications of the valon model to low-$p_T$ hadronic reactions
that are not perturbative. Such distributions are needed when the
parton degrees of freedom are released. If in some nuclear
reactions at some energy where the partons do not exhibit their
dynamical effect beyond the valons, as suggested by
Cs\"{o}rg\H{o} \cite{tc}  for heavy-ion collisions at SPS, then the
valon distributions are all that is needed. Since the exclusive
valon distribution is the absolute square of the wave function of
the proton in the valon representation, it is also the
recombination function of valons in forming a proton \cite{hvm}.
Thus the new values of $\alpha$ and $\beta$ found here affect
the calculation of hadron production at low $p_T$.

The valon model can, of course, also be applied to other hadrons
beside the proton. Although the  data on the PDF's of
the pion or the kaon are not of the same quality  as those of the
proton, some  data on the Drell-Yan and prompt photon
production initiated by mesons do exist. It will be a natural
extension of this work to determine the valon distributions in the
mesons by using the PDF's obtained from such data. By virtue of
the universality of the parton distributions in valons, what we
have found here from the proton is good enough for the mesons
also.

The affirmation of the validity of the valon model makes possible a
logical link between the bound-state problem of the hadrons in
terms of the constituent quarks and the scattering problem in
terms of the partons. The relationship between the wave functions
of the constituent quarks and the valon distributions was studied
in the context of form factors \cite{hl}. In view of the new
distributions  determined here, that problem needs to be revisited.
On the whole our understanding of the hadron structure problem
is enhanced by our  study here of the modern parton distribution
functions in the framework of the valon model.

\section*{Acknowledgment}     We are grateful to D.\ Soper for
helpful comments.  This work was supported, in part,  by the U.\
S.\ Department of Energy under Grant No. DE-FG03-96ER40972.

\begin{figure}[tbh]
\includegraphics[width=0.5\textwidth]{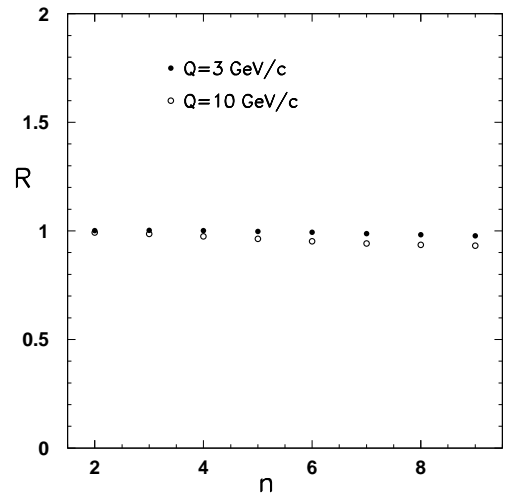}
\caption{The ratio of ratios defined in Eq.\,(\ref{3.9}).}
\end{figure}

\begin{figure}[tbh]
\includegraphics[width=0.5\textwidth]{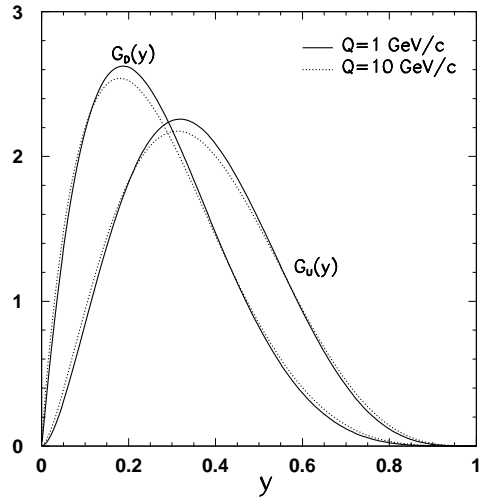}
\caption{The $U$ and $D$ valon distributions at two $Q$ values.}
\end{figure}

\begin{figure}[tbh]
\includegraphics[width=0.5\textwidth]{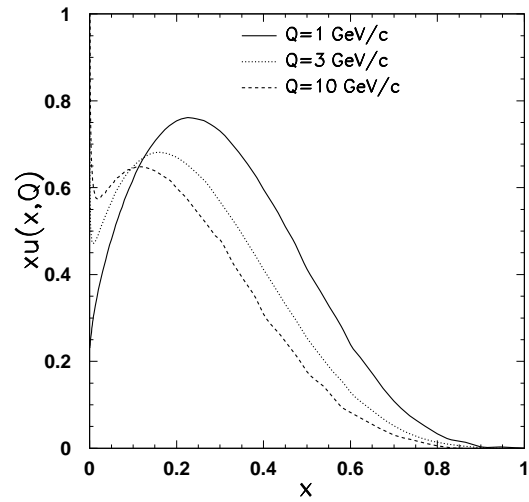}
\caption{The $u$ quark distribution functions from Ref.\cite{ct4}.}
\end{figure}

\begin{figure}[tbh]
\includegraphics[width=0.5\textwidth]{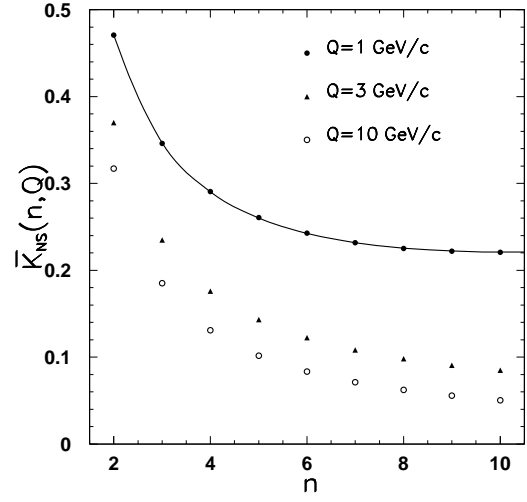}
\caption{The moments $\tilde K_{NS}$ for three values of
$Q$. The solid line is a fit by Eq.\,(\ref{4.9}).}
\end{figure}

\begin{figure}[tbh]
\includegraphics[width=0.5\textwidth]{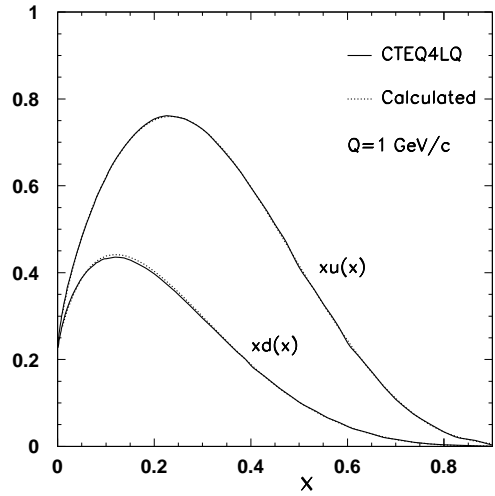}
\caption{The $u$ and $d$ quark distribution functions at
$Q=1$ GeV/c. The solid lines are from CTEQ \cite{ct4}; the dotted
lines are calculated in the valon model.}
\end{figure}

\begin{figure}[tbh]
\includegraphics[width=0.5\textwidth]{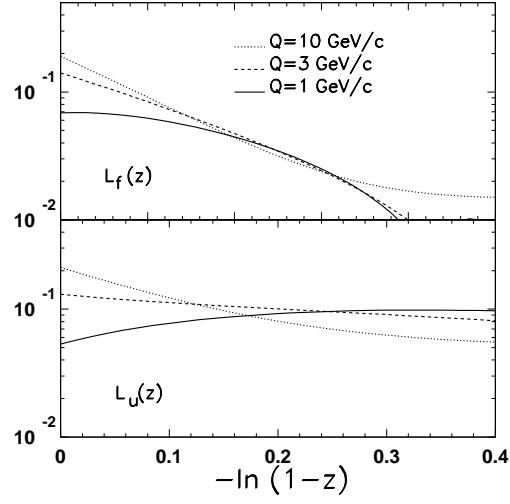}
\caption{The favored and unfavored sea quark
distributions in the valon at three values of $Q$.}
\end{figure}

\begin{figure}[tbh]
\includegraphics[width=0.5\textwidth]{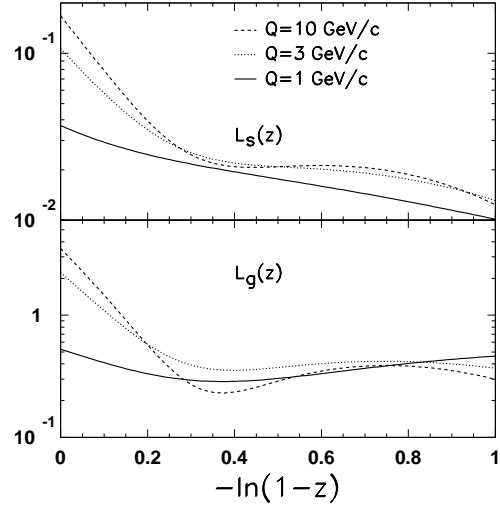}
\caption{The strange  quark and gluon
distributions in the valon at three values of $Q$.}
\end{figure}

\begin{figure}[tbh]
\includegraphics[width=0.5\textwidth]{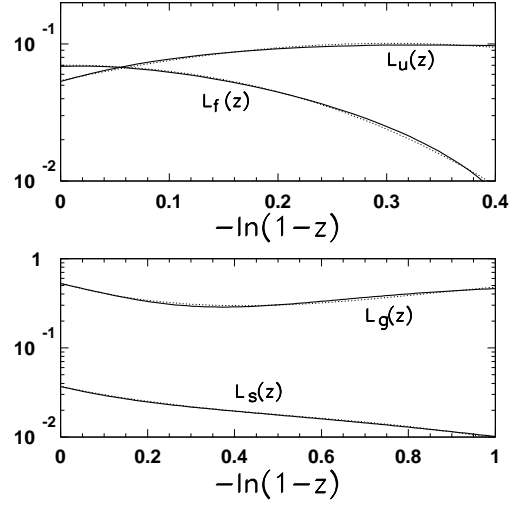}
\caption{Fits of the $f, u, s, g$ distributions at $Q=1$
GeV/c, as shown by the dotted lines.}
\end{figure}

\begin{figure}[tbh]
\includegraphics[width=0.5\textwidth]{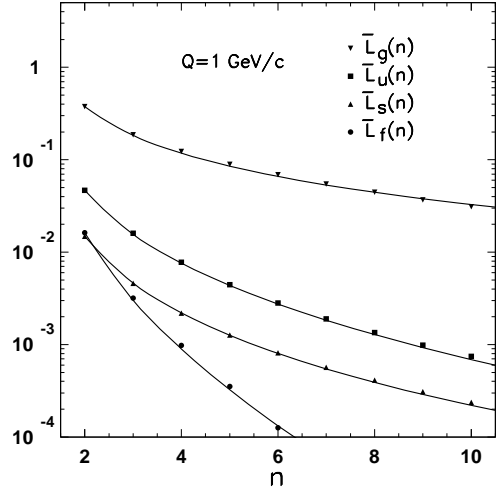}
\caption{The moments $\tilde L_g(n), \tilde L_u(n), \tilde
L_s(n), {\rm and} \tilde L_f(n)$ at $Q=1$ GeV/c and their fits.}
\end{figure}

\end{document}